\documentclass[conference]{IEEEtran}
\IEEEoverridecommandlockouts
\usepackage{cite}
\usepackage{verbatim}
\usepackage{amsmath,amssymb,amsfonts}
\usepackage{algorithmic}
\usepackage{graphicx}
\usepackage{textcomp}
\usepackage{xcolor}
\usepackage{svg}
\usepackage{siunitx}
\usepackage{bm}
\usepackage{hyperref}
\usepackage{pgfplots}
  \pgfplotsset{compat=newest}
  \usetikzlibrary{plotmarks}
  \usetikzlibrary{arrows.meta}
  \usepgfplotslibrary{patchplots}
  \usepackage{grffile}
\def\BibTeX{{\rm B\kern-.05em{\sc i\kern-.025em b}\kern-.08em
    T\kern-.1667em\lower.7ex\hbox{E}\kern-.125emX}}

\newcommand{\subd}{{d}}
\newcommand{\subq}{{q}}

\newcommand{\subs}{\textup{s}}

\begin{document}
\title{Moving-Horizon State Estimation for Power Networks and Synchronous Generators\\

\thanks{Research supported by the Swiss National Science Foundation under NCCR Automation, grant agreement 51NF40\_180545}
}

\author{\IEEEauthorblockN{Milos Katanic}
\IEEEauthorblockA{\textit{Power Systems Laboratory} \\
\textit{ETH Zurich}\\
mkatanic@ethz.ch}
\and
\IEEEauthorblockN{John Lygeros}
\IEEEauthorblockA{\textit{Automatic Control Laboratory} \\
\textit{ETH Zurich}\\
jlygeros@ethz.ch}
\and
\IEEEauthorblockN{Gabriela Hug}
\IEEEauthorblockA{\textit{Power Systems Laboratory} \\
\textit{ETH Zurich}\\
hug@eeh.ee.ethz.ch}
}

\maketitle

\begin{abstract}
Power network and generators state estimation are usually tackled as separate problems. We propose a dynamic scheme for the simultaneous estimation of the network and the generator states. The estimation is formulated as an optimization problem on a moving-horizon of past observations. The framework is a generalization of static state estimation; it can handle incomplete model knowledge and does not require static network observability by PMUs. The numerical results show an improved estimation accuracy compared to static state estimation. Moreover, accurate estimation of the internal states of generators without PMUs on their terminals can be achieved. Finally, we highlight the capability of the proposed estimator to detect and identify bad data.    
\end{abstract}


\section{Introduction}
\label{SEC_introduction}
State Estimation (SE) is an online monitoring tool indispensable for the secure and efficient operation of power systems \cite{abur}. Typically, SE relies on the measured data acquired through the Supervisory Control and Data Acquisition system, which has slow scan rates and no time stamps. Hence, SE is performed in a static manner using only the most recent set of measurements. The computed snapshot of the system state provides sufficient information for performing important functions, such as contingency analysis, power flow calculation, and voltage stability assessment for conventional power systems. The reason is that the conventional power systems are predominantly operated in a quasi-static regime, with changes driven by the slow changes in electricity consumption.\par

In recent years, electric power systems have been experiencing a shift from conventional energy sources to stochastic, power electronics-interfaced renewable energy sources \cite{milano}. Additionally, new types of loads are being introduced, such as electric vehicles, heat pumps, etc. The high penetration of renewables, alongside their intermittency, renders operating points less predictable and power system dynamics faster. Therefore, it is expected that real-time state estimates at sufficiently fast rates are going to be needed to achieve a reliable and efficient operation of the future power systems \cite{roles}. This observation has triggered a substantial research interest in dynamic SE (DSE) for power systems.\par 

DSE uses highly accurate, synchronized measurements from phasor measurement units (PMUs) to provide continuous information about the system states. In general, DSE for power systems can be performed in a centralized or decentralized manner \cite{motivation}. The former is achieved by sending measurement data to a central location and performing SE for the whole system \cite{sakis}, whereas the latter uses only local PMU data from a generator's terminals to estimate the internal states of the generator \cite{ukf}. Most of the published works on DSE employ recursive algorithms, such as the extended Kalman filter \cite{ekf}, the unscented Kalman filter \cite{ukf}, or particle filters \cite{pf}. Recently, \cite{armin} and \cite{mhe_pse} proposed moving-horizon estimation (MHE) schemes for DSE. \par

In this work, the centralized approach for DSE is adopted, as this approach offers the possibility to simultaneously estimate the state of the network in the form of nodal voltages as well as the internal states of the generators. In the literature, when performing centralized DSE, authors usually reduce the network to generator nodes, as in \cite{c_dse}, \cite{c_dse2}. This approach can only be applied if the models and parameters of all system components are known \cite{motivation}. Methods from \cite{sakis} and \cite{zhao} also assume the knowledge of the models of all system components. However, this is not a realistic assumption, especially with respect to the loads and stochastic generation. Hence, we refrain from carrying out network reduction and allow for incomplete model knowledge. Furthermore, approaches usually rely on the PMU measurements from generator terminals to perform DSE and to detect bad data. \par
As large-scale power systems include many heterogeneous components, obtaining accurate models of all of them may be very difficult. We propose a moving-horizon-based DSE that does not require the complete power systems model nor access to the generator PMUs. MHE is an optimization-based estimation technique that uses a sliding window of past observations to compute the state estimates. In the proposed optimization framework, both the internal states of generators and nodal voltages are decision variables; hence, we can estimate the complete system state without requiring static network observability by PMU data. Even though the present work concentrates on conventional generation, the method can be extended to include the renewable generation. The MHE framework also offers the flexibility to handle nonlinearity and state constraints explicitly \cite{mhe_raw}. In the context of power systems, this allows one to explicitly include network algebraic constraints into the estimation framework. We show that this approach is, in fact, a generalization of SSE. Hence, the largest normalized residual (LNR) test can be applied to detect and identify bad data. \par

The remainder of the paper is organized as follows. In Section \ref{SEC_modeling}, we derive the power system model. Section \ref{SEC_mhe} presents the estimation framework along with the bad data rejection scheme. In Section \ref{SEC_results}, simulation results are discussed. Finally, Section \ref{SEC_conclusion} concludes the paper.
\section{Power System Model}
\label{SEC_modeling}
We consider a power system comprising synchronous generators (SGs), loads, and a network interconnecting them. Phasor representation can be used to capture electromechanical transients, whereas fast, electromagnetic transients are neglected \cite{machkowski}. Let $n$ be the total number of nodes in the network, $\mathcal{G}$ the set of nodes containing SGs with known models, and $\mathcal{Z}$ the set of all nodes with zero injections, which includes:
\begin{itemize}
    \item Nodes with no load and no generation;
    \item Nodes with loads that can be accurately modeled by a specified complex admittance. 
\end{itemize}
If the model of a load is unknown, the respective node is not included in $\mathcal{Z}$. If a load can be modeled as a complex admittance, it is integrated into the circuit model of the grid. For each node $i$, let $\mathcal{N}_i$ be the set of neighboring nodes.
\subsection{Synchronous Generator Model}
Synchronous generators constitute the majority of commercial electrical energy production. Of the several mathematical models of SGs that have been proposed in the literature, here we employ the fourth-order model in the local $dq$ reference frame \cite{Sauer1998}. Therefore, the model of a SG connected to node $i$ is given by:
\begin{subequations}
\label{SGmodel}
\begin{align}
    \dot{\delta}_i &= \omega_{n}\, \Delta \omega_i,\\
    \Delta \dot{\omega}_i &= \frac{1}{2H_i} \left(p_{m,i} - p_{e,i}  - D_i \Delta \omega_i\right),\\
    \dot{e}'_{d,i} &=\frac{1}{T'_{q0,i}} \left( - e'_{d,i} + (x_{q,i} - x'_{q,i})i_{q,i} \right),\\
    \dot{e}'_{q,i} &=\frac{1}{T'_{d0,i}} \left(E_{\textup{fd},i} - e'_{d,i} - (x_{d,i} - x'_{d,i})i_{d,i} \right),
\end{align}
\end{subequations}
where $\delta_i$ is the rotor angle, $\omega_{{n}}$ is the rated rotor angular frequency, $\Delta \omega_i$ is the relative rotor speed deviation, i.e., $({\omega_{r,i} - \omega_n})/{\omega_n}$, $H_i$ is the inertia constant, $p_{m,i}$ is the mechanical power, $p_{e,i}$ is the electric power provided by the machine, $D_i$ is the damping coefficient, $E_{\textup{fd},i}$ is the internal field voltage, $e'_{d,i}$ and $e'_{q,i}$ are the voltages behind the transient reactances, $T'_{d0,i}$ and $T'_{q0,i}$ are the open-circuit time constants, $x_{d,i}$ and $x_{q,i}$ are the synchronous reactances, $x'_{\subd,i}$ and $x'_{\subq,i}$ are the transient reactances in the $d$ and $q$ axes, respectively. Model (\ref{SGmodel}) is given in per-unit values, i.e., the quantities are expressed relative to their base values. \par
Furthermore, we consider the IEEE DC1A automatic voltage regulator \cite{Sauer1998} without power system stabilizer given by:
\begin{subequations}
\begin{align}
    T_{E,i} \dot{E}_{\textup{fd},i}& = -K_{E,i}E_{\textup{fd},i} + V_{R,i},\\
    T_{F,i} \dot{R}_{f,i} &= - R_{f,i} + \frac{K_{F,i}}{T_{F,i}}E_{\textup{fd},i},\\
    T_{A,i}\dot{V}_{R,i} &= -V_{R,i} + K_{A,i}R_{f,i} - \frac{K_{A,i}K_{F,i}}{T_{F,i}}E_{\textup{fd},i} \notag \\
    &\quad+ K_{A,i}(v_{\textup{ref},i} - v_i),
\end{align}
\end{subequations}
and the simple TGOV1 turbine model \cite{governor} given by:
\begin{subequations}
\label{TGmodel}
\begin{align}
    T_{1,i}\dot{p}_{\textup{sv},i} &= p_{\textup{ref},i}-\frac{\Delta \omega_i}{R_i} - p_{\textup{sv},i},\\
    T_{3,i} \dot{p}_{m,i} &= \frac{T_{2,i}}{T_{1,i}} (p_{\textup{ref},i} - \frac{\Delta \omega_i}{R_i} - p_{\textup{sv},i}) + p_{\textup{sv},i} - p_{m,i}.
\end{align}
\end{subequations}
Here, $T_{E,i}$ is the exciter time constant, $K_{E,i}$ is the exciter field constant without saturation, $v_{R,i}$ is the pilot exciter voltage, $T_{F,i}$ and $K_{F,i}$ are the stabilizer gain and time constant, respectively, $R_{f,i}$ is the feedback rate, $K_{A,i}$ and $T_{A,i}$ are the voltage regulator gain and time constant, respectively, $v_i$ is the terminal voltage magnitude, $T_{1,i}$ is the steam bowl time constant, $p_{\textup{sv},i}$ is the steam valve position, $T_{2,i}$ and $T_{3,i}$ are the turbine lead and lag time constant, and $R_i$ is the generator droop. Generator set points for the active power $p_{\textup{ref},i}$ and the voltage magnitude $v_{\textup{ref},i}$ are assumed to be known. The power $p_{\textup{ref},i}$ is obtained from the economic dispatch and the voltage $v_{\textup{ref},i}$ is set by the operator.  \par
The state of the SG model connected to node $i$ from (\ref{SGmodel})--(\ref{TGmodel}) can be defined as:\\
$\bm{x}_i = 
    \begin{bmatrix}
    \delta_i, \Delta \omega_i, e_{\subd,i}', e_{\subq,i}', p_{\textup{sv},i}, p_{m,i}, E_{\textup{fd},i}, R_{f,i}, V_{R,i}
    \end{bmatrix}^\top, i \in \mathcal{G}.
$
The electric power is given by
\begin{equation}
    p_{e,i} = e_{q,i}'i_{q,i} + e_{d,i}'i_{d,i} + (x_{d,i}' - x_{q,i}')i_{d,i}i_{q,i}.
\end{equation}
Similarly to decentralized DSE \cite{anag}, we can express the output current of the generator as a function of its internal states and the terminal voltage phasor $(v_i, \theta_i)$ according to
\begin{equation}
    \begin{bmatrix}
    i_{\subd,i} \\
    i_{\subq,i} \\
    \end{bmatrix} = \begin{bmatrix}
    r_{\subs,i}  & x_{\subq,i}'\\ -x_{\subd,i}' & r_{\subs,i} 
    \end{bmatrix}^{-1}\begin{bmatrix}
    v_i\sin (\theta_i - \delta_i) + e_{\subd,i}' \\ -v_i\cos (\theta_i - \delta_i) + e_{\subq,i}'
    \end{bmatrix},
    \label{inj1}
\end{equation}
which can be further transferred to a common rotating coordinate system by
\begin{equation}
\begin{bmatrix}
    i_{D,i}\\
    i_{Q,i}
    \end{bmatrix} = \frac{S_{n,i}}{S_{b}}\begin{bmatrix}
    \sin \delta_i &\cos \delta_i\\
    -\cos \delta_i& \sin \delta_i
    \end{bmatrix}
    \begin{bmatrix}
    i_{d,i}\\
    i_{q,i}
    \end{bmatrix}.
    \label{inj2}
\end{equation} 
The positive factor $\frac{S_{n,i}}{S_\textup{b}}$ scales the values from each generator to the system per-unit values, where $S_n^i$ is the rated power of the SG at node $i$ and $S_b$ is the system base power. The matrix in (\ref{inj1}) is invertible because the SG inductances and resistances are positive quantities. Eq. (\ref{inj1}) and (\ref{inj2}) can be combined and compactly written as
\begin{equation}
    \bm{i}_i = \bm{g}_{g} (\bm{x}_i, v_i, \theta_i), \quad i \in \mathcal{G},
    \label{inj3}
\end{equation}
where subscript $i$ denotes the respective node. \par
For notational purposes, we collect the vectors of the states of individual generators into a single vector $\bm{x}$ containing all differential states in the system.
Similarly, the vectors of the network voltage magnitudes and phases can be denoted by $\bm{v}$ and $\bm{\theta}$.
In order to implement the derived model, the equations need to be discretized. The discrete-time state-space representation from (\ref{SGmodel})--(\ref{inj1}) can be obtained by the forward Euler method and is given by
\begin{equation}
    \bm{x}_{i,k+1} = \bm{f}\left(\bm{x}_{i,k}, {v}_{i,k}, \theta_{i,k}\right) + \bm{\omega}_{i,k}, \quad i \in \mathcal{G},
    \label{Euler}
\end{equation}
where $\bm{\omega}_{i,k}$ is the additive noise due to the discretization errors and the model mismatch. Subscript $k$ denotes the time instant.

\subsection{Network Model}
\label{SUBSEC_network}
We model the network with algebraic equations as the dynamics of transmission networks are much faster than the internal dynamics of the rotating machines \cite{machkowski}. The network consists of interconnected transmission lines and transformers, which can be modeled as equivalent circuits. The nodal current injections in $D$ and $Q$ axes following Kirchoff's current law at node $i$ are given by
\begin{align}
    &i_{{D},i}^\textup{inj} =  \sum_{j \in \mathcal{N}_i} \Big[\left( v_i\cos \theta_i - v_j \cos \theta_j \right)g_{ij} - \notag \\ 
    &\left( v_i\sin \theta_i - v_j \sin \theta_j \right)b_{ij}\Big] + v_i \cos \theta_i g^{\textup{sh}}_i-v_i \sin \theta_i b^{\textup{sh}}_i , 
    \label{i_d}
\end{align}
\begin{align}
    &i_{Q, i}^{\textup{inj}} = \sum_{j \in \mathcal{N}_i} \Big[\left( v_i\cos \theta_i - v_j \cos \theta_j \right)b_{ij} + \notag\\
    & \left( v_i\sin \theta_i - v_j \sin \theta_j \right)g_{ij}\Big] + v_i \sin \theta_i g_{i}^\textup{sh}+v_i \cos \theta_i b_{i}^\textup{sh},
    \label{i_q}
\end{align}
where $g_{i}^\textup{sh}$ and $b_{i}^\textup{sh}$ are the conductance and susceptance of the shunt branch connected to bus $i$, and $g_{ij}$ and $b_{ij}$ are the admittance and susceptance of the series branch connecting buses $i$ and $j$, respectively. For the sake of notation and assuming network parameters known, (\ref{i_d}) and (\ref{i_q}) can be rewritten together as follows:
\begin{equation}
    \bm{i}^\textup{inj}_i = \bm{g}_{n,i} \left(\bm{v}, \bm{\theta} \right), \quad i=1,\ldots,n,
    \label{network_current}
\end{equation}
where vectors $\bm{v}$ and $\bm{\theta}$ constitute the state of the network.

\subsection{Synchronous Generators and Network Coupling}
\label{coupling}
The states of the SGs and the network can be linked at each time instant $k$ via the injected currents. Hence, for $i \in \mathcal{G}$, the generator output current from (\ref{inj3}) and the network injected current from (\ref{network_current}) can be set equal as
\begin{align}
    \bm{g}_{g}\left(\bm{x}_{i,k}, {v}_{i,k}, {\theta}_{i,k} \right) + \bm{e}_{i,k} &= \bm{g}_{n,i} \left ( \bm{v}_k, \bm{\theta}_k \right),   \quad  i \in \mathcal{G},
    \label{current_deviation}
\end{align}
where $\bm{e}_{i,k}$ is the noise term associated with the SG model mismatch. \par
For zero injection nodes, (\ref{network_current}) simplifies to
\begin{equation}
    \bm{0} = \bm{g}_{n,i}\left( \bm{v}_k, \bm{\theta}_k \right), \quad  i \in \mathcal{Z}.
    \label{zero_nodes}
\end{equation}
There is no noise in (\ref{zero_nodes}), because we assume that the network model and parameters are known with sufficient accuracy. Since the network parameters are needed for SSE, these parameters are already available to transmission system operators. Recall that the loads that can be accurately modeled with a specified conductance and susceptance are included in $\mathcal{Z}$, and their injected current is zero. \par
For the nodes with unknown components or interface nodes to parts of the network that are out of interest, the injected current is unknown. We can simply omit the corresponding equation in (\ref{network_current}) from further consideration; no forecasting methods are employed to predict the injection as these approaches may be unreliable under sudden changes.  
\subsection{Measurement Functions}
DSE relies on the synchronization and high sampling rate of PMU measurements, which use the global positioning system to obtain an accurate time reference. Their output includes the magnitude and the phase of the complex phasor representing a sinusoidal signal \cite{phadke}. The measured signal can either be the voltage or the current, depending on the type of measurement. To use these measurements for SE, they must be linked to the system states. As the algebraic voltage variables are part of the system state, the measurement function is given the same way as for SSE. Hence, discrete-time voltage, current flow, and current injection measurements can be expressed as functions of nodal voltages:
\begin{equation}
    \bm{y}_k = \bm{h}^\textup{PMU}(\bm{v}_k, \bm{\theta}_k) + \bm{e}^\textup{PMU}_k.
    \label{PMU_function}
\end{equation}
Here, $\bm{y}_k$ denotes the vector of PMU data (phase and magnitude of the measured signals), $\bm{h}^\textup{PMU}$ denotes the measurement function, and $\bm{e}^\textup{PMU}_k$ is the measurement noise. We neglect the remote terminal unit measurements because of their slow update rate and the absence of time stamps.

\section{Moving-Horizon Estimation Method}
\label{SEC_mhe}
The previously derived power system model and the PMU measurement functions are now incorporated into the moving-horizon estimation framework.
\subsection{Optimization Problem Formulation}
The estimation time horizon at each time step $t$ is divided uniformly from $s$ to $t$ by $L = t - s +1$ sampling points. As already mentioned, we propose the following optimization problem to simultaneously estimate the internal states of generators and nodal voltage phasors: 
\begin{equation}
\min_{{\{ \bm{x}_k , \bm{\theta}_k , \bm{v}_k \}_{k=t-L+1}^{t}}} \bigg( \lVert \bm{x}_{t-L+1} - \bm{\bar{x}}_{t-L+1}\rVert_{\bm{P}_0}^2 + \!\! \sum_{k = t-L+1}^t \! \lVert \bm{e}^\textup{PMU}_k \rVert^2_{\bm{R}}  \notag\\
\end{equation}
\begin{subequations}
\label{opt}
\begin{equation}
\label{mhe_opt}
+\sum_{i \in \mathcal{G}} \sum_{k=t-L+1}^{t-1} \lVert \bm{\omega}_{i,k} \rVert^2_{\bm{P}}  +   \sum_{i \in \mathcal{G}} \sum_{k=t-L+1}^{t} \lVert \bm{e}_{i,k} \rVert^2_{\bm{Q}} \bigg)
\end{equation}
\begin{equation}
\textrm{s.t.}\quad \bm{g}_{n,i} \left(\bm{v}_k, \bm{\theta}_k \right) = \bm{0}, \, \, k = t-L+1,\ldots,t;\, \,i \in \mathcal{Z}, 
\label{mhe_opt_constr}
\end{equation}
\end{subequations}
\noindent
where $\lVert \bm{a} \lVert^2_{\bm{A}} = \bm{a}^\top \bm{A} \bm{a}$; the first term in (\ref{mhe_opt}) represents the arrival cost; the second, third and fourth are penalty terms on $e^\textrm{PMU}_k$ in (\ref{PMU_function}), $\omega_{i,k}$ in (\ref{Euler}), and $e_{i,k}$ in (\ref{current_deviation}), respectively; subscripts $k$ and $i$ denote the time instant and node index, respectively. Weighting matrices $\bm{P}_0$, $\bm{R}$, $\bm{P}$, $\bm{Q}$ are diagonal and positive definite and correspond to tuning parameters. $\bm{P}_0$ is the fixed weighting matrix for the arrival cost penalization. Arrival cost can be seen as an implicit inclusion of past measurement data, not explicitly included in the cost function. $\bm{\bar{x}}_{t-L+1}$ is the state estimate at time $t-L+1$ obtained from the solution of the optimization problem at time instant $t-1$. This term is initialized with the best known guess of the system state at the initial time instant. The length of the estimation horizon $L$ is a designer's choice and represents a trade-off between performance and computational time. The quadratic cost function was chosen because its minimization under Gaussian noise can be interpreted as the maximum-likelihood estimator.
The solution of (\ref{opt}) contains the estimates of the previous $L$ system states. The most recent estimate, $\bm{x}_t, \bm{u}_t, \bm{\theta}_t$, is taken as the current state of the power system; the other estimates (except for the oldest one) are used to warm-start the optimization at the following time instant. \par 
Note that if all SG models are unknown and $L=1$, problem (\ref{opt}) reduces to SSE. Hence, SSE can be considered a special case of this approach. Of course, SSE can only output the unique solution if the network is statically observable. Here, we give an intuition why the proposed dynamic estimator requires fewer PMU measurements than SSE; rigorous analysis will be part of future work. Compared to SSE, each SG model adds another nine (model order) unknown states to the problem, but at the same time generates eleven additional equations (nine dynamic evolution equations and two current injection equations) for each time instant in the horizon. Hence, an additional measurement redundancy is created with each included dynamic model of an injector (load or generator). The intuition also tells us that parts of the network with more unknown injectors require more PMU measurements to guarantee the uniqueness of the solution.  
\subsection{Solution Framework}
The optimization problem (\ref{opt}) is a constrained nonlinear weighted least squares problem. Hence, it can be written as
\begin{subequations}
\label{residual_function}
\begin{align}
    \min_{\bm{X}} & \quad \lVert  \bm{h}(\bm{X}) \rVert^2_{\bm{W}}\\
    \textrm{s.t.}& \quad \bm{c}(\bm{X}) = 0. 
\end{align}
\end{subequations}
Here, $\bm{X}$ represents the vector of all differential and algebraic states, function $\bm{h}$ is the nonlinear residual function containing all residual terms from (\ref{mhe_opt}), $\bm{W}$ is the diagonal matrix generated from the weighting matrices of the individual least squares terms, and finally, $\bm{c}$ is the constraint function (\ref{mhe_opt_constr}). Recall that SSE is also usually formulated as a weighted least squares problem \cite{abur}. This fact allows us to leverage some already well-established techniques from SSE, such as the Gauss-Newton method for finding a local minimum and the LNR test for bad data detection.   \par
The minimiser of (\ref{residual_function}) can be computed iteratively using the constrained Gauss-Newton method. The next iterate, $\bm{X}^{k+1}$, is computed by solving the following set of linear equations \cite{abur}:
\begin{equation}
    \label{GN}
    \begin{bmatrix}
    \bm{H}^\top \bm{W}\bm{H} & \bm{C}^\top\\
    \bm{C} & \bm{0}
    \end{bmatrix}  \begin{bmatrix}
    \Delta \bm{X}\\
    -\bm{\lambda}
    \end{bmatrix} = \begin{bmatrix}
    \bm{H}^\top \bm{W}\bm{h}(\bm{X}^k)\\
    -\bm{c}(\bm{X}^k)
    \end{bmatrix}.
\end{equation}
Here, $
\bm{H}= \frac{\partial \bm{h}\left(\bm{X}\right)}{\partial \bm{X}}\Bigr|_{\substack{\bm{X}=\bm{X}^{k}}},
\bm{C}= \frac{\partial \bm{c}(\bm{X})}{\partial \bm{X}}\Bigr|_{\substack{\bm{X}=\bm{X}^{k}}}  
$
are the Jacobians of the residual function and the constraint function at the current iterate, respectively, $\bm{\lambda}$ is the vector of Lagrange multipliers, and $\Delta \bm{X} = \bm{X}^{k+1} - \bm{X}^k$. 

\subsection{Bad Data Detection}
Gross errors in PMU measurements can appear due to various reasons, such as communication issues, sensor malfunction, and cyber-attacks, among others. To prevent these outliers from degrading the performance of SE, the state estimator can be equipped with a bad data processing scheme. Bad data detection and identification for SSE has been addressed thoroughly in the literature. The most commonly used post-processing step is the LNR test \cite{abur}, which removes or compensates measurements whose normalized residuals exceed a statistical threshold. The normalized measurement residuals are calculated according to
$
    r_{j}^N = \frac{\lvert r_{j} \rvert }{\sqrt{{\Omega}_{jj}}},
$
where $r_{j}$ is the $j$\textsuperscript{th} raw residual and $\Omega_{jj}$ is the $j$\textsuperscript{th} diagonal entry of the residual covariance matrix. The residual covariance matrix for the equality constrained state estimation can be calculated according to \cite{eq_res}
$
    \bm{\Omega} = \bm{W}^{-1} - \bm{H}\bm{E}_\textup{ul}\bm{H}^\top,
$
where $\bm{E}_\textup{ul}$ is the upper left corner of the inverse of the coefficient matrix from (\ref{GN}), i.e.,
$
    \begin{bmatrix}
    \bm{E}_\textup{ul} & \bm{E}_\textup{ur}\\
    \bm{E}_\textup{ll} & \bm{E}_\textup{lr}
    \end{bmatrix} = \begin{bmatrix}
    \bm{H}^\top \bm{W}\bm{H} & \bm{C}^\top\\
    \bm{C} & \bm{0}
    \end{bmatrix}^{-1}.
$
After removing the measurement corresponding to the largest normalized residual, the SE is rerun, and the whole procedure is repeated until there are no bad data in the measurement set.  
\section{Simulation Results}
\label{SEC_results}

\begin{figure}
    \centering
    \includegraphics[width = 3.5in]{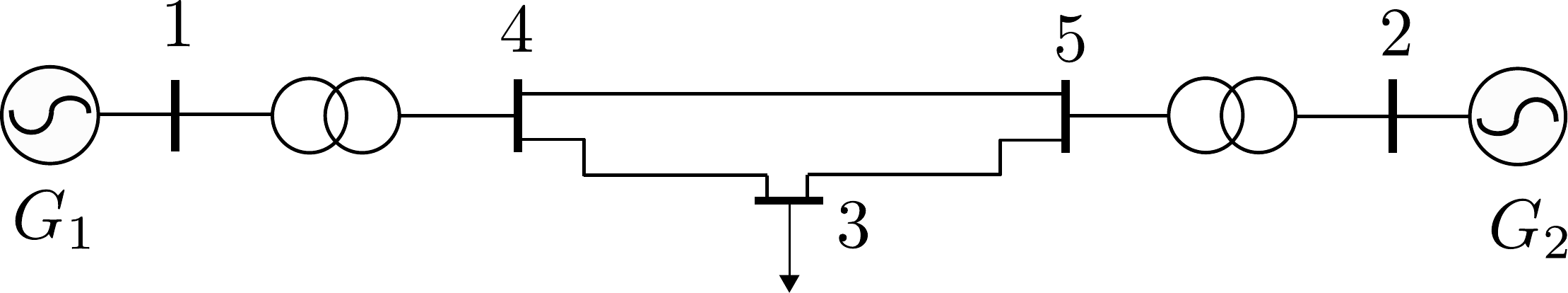}
    \caption{Topology of the five-bus, two-generator power system.}
    \label{five_bus}
\end{figure}

In this section, we present numerical results obtained on the small-scale, five-bus, two-generator, and one-load power system shown in Fig. \ref{five_bus}. 
The goal is to test the performance of the estimation scheme presented in Section \ref{SEC_mhe}, here denoted by MHE, and compare it to SSE. In real-world applications, models used for estimation are only an approximate description of the real system behavior. To this end, we intentionally use higher-order synchronous machine models for simulating the system's behaviour. The employed simulation model is the Simulink subtransient model that also includes stator dynamics. Reduced order parameters of SGs and network parameters are considered to be known. Furthermore, we consider the model of the load at node 3 to be unknown. The PMU measurements necessary for the estimation were generated by adding white Gaussian noise with the variance $\sigma^\textup{PMU}=10^{-6}$ to the simulation results for both the magnitude and the phase of the current and voltage measurements. We assume that all PMU measurements deployed in the network have reporting rates $F_\textup{s} = \SI{100} {Hz}$ \cite{pmu_standard} and are available without time delay. The chosen length of the estimation horizon was set to $L=3$ with the same time resolution as the PMU measurements. Longer horizons did not show significant accuracy improvement on the tested scenarios. In total, vector $X$ contains 84 states. All simulations are implemented in Matlab.

\subsection{Test Case 1}
For the first scenario, we consider the system in Fig. \ref{five_bus} with only one PMU device. The PMU measures the voltage at node 4 and the current between nodes 4 and 5. With the given measurements, the network is statically unobservable. The disturbance is characterized by a step change in the active and reactive load power consumption at $t=3.5$ s. Note that this information is not available to the estimator, as the load model is considered to be unknown.\par 
\begin{figure}
    \centering
    \resizebox{\columnwidth}{!}{
    \input{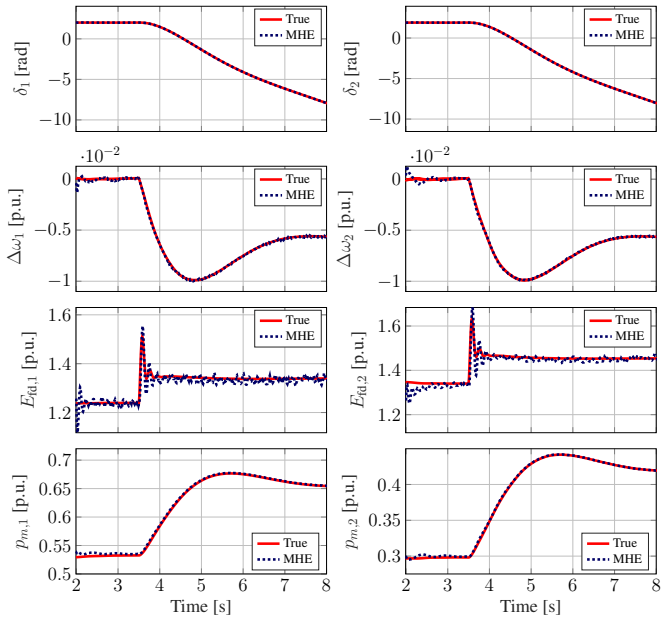}}
    \caption{Results for test case 1. Estimation of the internal states of SGs: SG1 on the left and SG2 on the right. From top to bottom: the rotor angle, rotor speed, internal field voltage, and mechanical power are shown. }
    \label{UKF_vs_MHE}
\end{figure}
The algorithm for the MHE converged on average after $2$ iterations. Fig. \ref{UKF_vs_MHE} shows the accuracy of the estimation of the dynamic states of the generators: rotor angle, rotor speed, internal field voltage, and mechanical power. The estimator exhibits very good tracking performance, with the estimation error of all the estimates close to zero at all times. This result indicates that even without access to SG terminal PMU measurements, accurate state estimation of generator states can be achieved. In addition, MHE estimates the algebraic voltages in the network. 
\begin{figure}
    \centering
    \resizebox{\columnwidth}{!}{
    \input{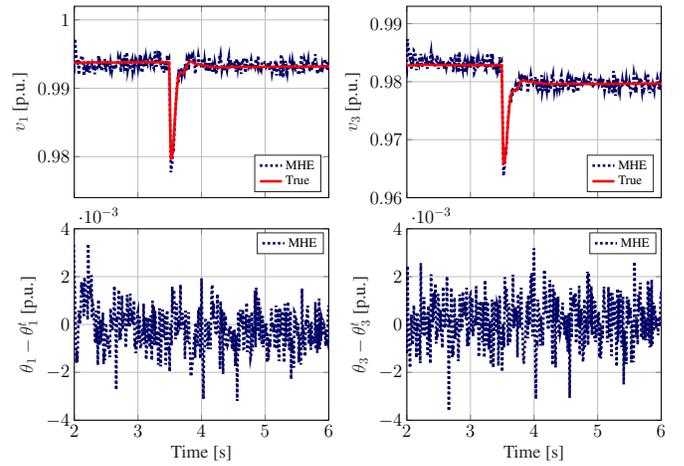}}
    \caption{Results for test case 1. Algebraic state estimation: node 1 on the left and node 3 on the right. The upper graphs show the voltage magnitude; the lower graphs show the voltage phase error.}
    \label{MHE_V}
\end{figure}
The estimated and the true voltages of nodes 1 and 3 are shown in Fig. \ref{MHE_V}. The estimation error is close to zero during both steady-state and transient operation. The mean square error (MSE) is used to evaluate the estimation accuracy. It is given by
$
    \sigma_i = \frac{1}{2K} \sum_{k=1}^K \left[\left(\hat{v}_i(k) - {v}^t_i(k)\right)^2 +  \left(\hat{\theta}_i(k) - {\theta}^t_i(k)\right)^2\right], i=1,...,n,
$
where $\hat{}$ denotes the estimated quantities, superscript $t$ denotes the true quantities, and $K$ the number of the state estimation time instants. 
The MSE of the voltage estimation for nodes 1 to 5 is: $\sigma =10^{-6} \{1.062, 1.137, 1.140, 0.995, 0.998 \}$. It can be seen that all voltages in the network are accurately estimated. By contrast, the SSE problem is underdetermined. In conclusion, even with very few PMU measurements, an accurate, real-time estimation of all states in the power system can be achieved. 

\subsection{Test Case 2}
\begin{figure}
    \centering
    \resizebox{\columnwidth}{!}{
    \input{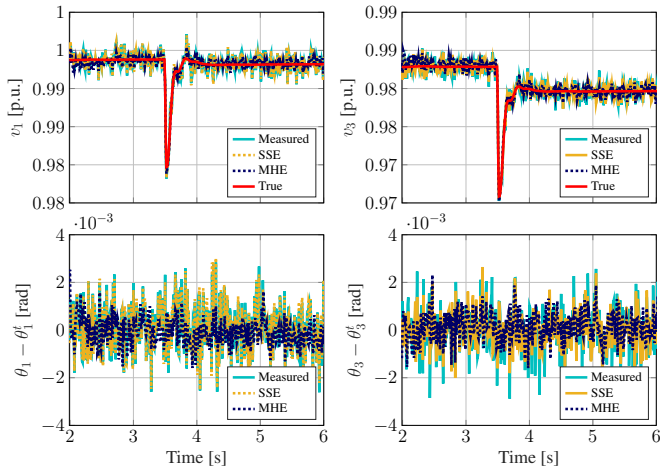}}
    \caption{Results for test case 2. Algebraic state estimation: node 1 on the left and node 3 on the right. The upper graphs show the voltage magnitude; the lower graphs show the voltage phase error.}
    \label{SSE_vs_MHE}
\end{figure}

To compare the performance of the proposed estimator with SSE, the network is made statically observable by PMU data in this test case. For this purpose, we use four PMU measurements. The voltage phasor is measured at nodes 1, 3, and 5. The current flow is measured between nodes 4 and 5. If we take into account two zero injections at nodes 4 and 5, the network is statically observable. Fig. \ref{SSE_vs_MHE} compares the estimation accuracy of the two approaches. In the top graph, the magnitude of the voltage is shown; at the bottom, the phase estimation error is shown. 
We can infer that the MHE displays higher accuracy of the voltage estimation compared to SSE. This conclusion is verified in Table \ref{tab:mhe_vs_sse}, where the estimation accuracy is quantitatively compared for all the nodes.
\begin{table}[]
    \caption{Mean square error of the MHE and SSE for test case 2.}
    \centering
    \begin{tabular}{|c|c c c c c|}
    \hline
    \textbf{Node} & 1& 2&3&4&5 \\
    \hline
    $\sigma^\textrm{MHE}_i [10^{-6}]$  &  0.311 & 0.405 & 0.395 & 0.338 & 0.342\\
    $\sigma^\textrm{SSE}_i [10^{-6}]$  &  0.977 & 0.990 & 0.650 &  0.380 & 0.381\\
    \hline 
    \end{tabular}
    \label{tab:mhe_vs_sse}
\end{table}
\subsection{Impact of Bad Data}
\begin{figure}
    \centering
    \resizebox{0.75\columnwidth}{!}{
    \input{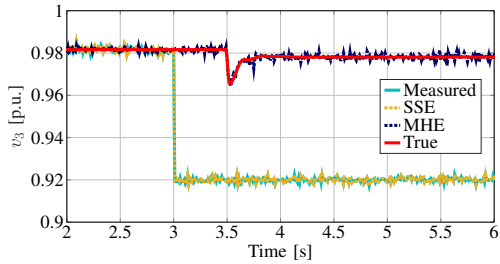}}
    \caption{Results for test case 3 including bad data. The voltage magnitude of node 3 is shown.}
    \label{SSE_vs_MHE_cs}
\end{figure}

To evaluate the impact of bad data on the proposed state estimation, we consider a case with three PMU measurements, deployed as follows: voltage measurements at nodes 3 and 4, and a current measurement between nodes 4 and 5. The bad data corresponding to the value of {0.92} p.u. are injected into the magnitude of $v_3$ starting at $t = \SI{3}{s}$. For SSE, this represents a case of bad data in a critical measurement, which cannot be detected by LNR. On the other hand, MHE creates additional measurement redundancy using dynamic equations. Fig. \ref{SSE_vs_MHE_cs} showcases the ability of the proposed MHE to detect and identify bad data in this case. 
The corrupted measurement was removed, and the performance of MHE was not impaired. 
\section{Conclusions and Future Work}
\label{SEC_conclusion}
We presented a moving-horizon-based method for simultaneous state estimation of power network and generators. The estimator does not require the placement of PMUs on generators' terminals and can handle missing model characteristics; it only requires accurate network model parameters. The performance of the proposed estimation technique was compared against the SSE. The results show that with each incorporated dynamic model of SGs, the estimation accuracy could be improved and the number of required PMUs reduced. The results highlight the capability of the proposed estimator to achieve an accurate state estimation even if the network is not statically observable by the available PMUs. Thus, our approach offers the flexibility to gradually include dynamic generator models into power systems state estimation. Furthermore, the post-processing LNR test can detect and identify bad data in PMU measurements. \par
Additional work is planned on including renewable energy sources in the tested scenarios and analyzing the effect of multiple correlated bad data and cyber-attacks.  

\bibliography{bibtex.bib}
\bibliographystyle{ieeetr}

\end{document}